\documentclass[prd,twocolumn,showpacs]{revtex4}
\usepackage{graphicx,amsmath,amssymb}
\newcommand{\haak}[1]{\left(#1\right)}

\renewcommand{\imath}{\text{i}}

\begin{document}

\title{Has DAMA Detected Self-Interacting Dark Matter?} 
\author{Saibal Mitra}
\affiliation{Instituut voor Theoretische Fysica,\\ Valckenierstraat 65,\\ 1018 XE Amsterdam, The Netherlands}

\date{\today}
\begin{abstract}
We consider a model in which dark matter consists of a large self-interacting component (SIDM particles) and a small component with strong interactions with baryons (SIMPs). We show that the DAMA annual modulation signal can be caused by SIDM particles interacting with SIMPs trapped in iodine nuclei in the DAMA/NaI detector.

\end{abstract}

\maketitle
The nature of dark matter (DM) remains a mystery, despite many years of efforts to understand it. It has been shown \cite{ost} that collisionless dark matter correctly predicts the large scale structure of the universe. It is less clear, however, if collisionless dark matter is consistent with observations on the galactic scale and smaller.
Some studies have suggested that there are discrepancies with observations. It has been found that:
\begin{itemize}
\item More satellite galaxies are predicted in simulations than are observed \cite{sat}.
\item Simulations of disk galaxies contain less mass and angular momentum than is observed \cite{disk}.
\item Simulated dark matter halos of galaxies have steeper density profiles compared to observations \cite{cusp}.
\end{itemize} 
Dark matter consisting of particles with strong self-interactions (SIDM particles) have been suggested as a solution to these problems \cite{stein,wand}. The cross section between SIDM particles, $\sigma_{\text{dd}}$, needed is in the range \cite{stein,wand}:
\begin{equation}\label{range}
8\times 10^{-25}\text{ cm}^{2}\frac{m_{\text{SIDM}}}{\text{GeV}}\lesssim\sigma_{\text{dd}}\lesssim
1\times 10^{-23}\text{ cm}^{2}\frac{m_{\text{SIDM}}}{\text{GeV}}. 
\end{equation}
Here $m_{\text{SIDM}}$ is the mass of the SIDM particle.
While some other studies have casted doubt on the self-interacting dark matter hypothesis and have suggested upper limits on $\sigma_{\text{dd}}$ of the order of $\sim 10^{-25}$ cm$^2$ $m_{\text{SIDM}}$/GeV, see e.g. \cite{dbt}, in a recent new study \cite{ahn} it is argued that $\sigma_{\text{dd}}$ could be much higher, about $3.6\times 10^{-22}$ cm$^2$ $m_{\text{SIDM}}$/GeV.

Cross sections between dark matter particles could thus be of the same order as for particles interacting with each other via the strong force. This has led some to speculate that DM particles could interact with both themselves and with baryons through the strong force \cite{wand}. DM particles interacting with baryons with such strong interactions had been postulated earlier by Dover, Gaisser, Steigman and Wolfram \cite{dov}, and have been named SIMPs (Strongly Interacting Massive Particles). SIMPs have been proposed as a solution to the ultra high energy cosmic ray problem \cite{chung} and as an explanation for the absence of cooling flows \cite{chuz}. Constraints on SIMP dark matter are given in \cite{wand, mcg, jav}.

It has been shown \cite{moh,tep} that SIMPs can bind to heavy nuclei, forming anomalously heavy isotopes. Searches for such isotopes have been performed in gold and iron recently \cite{jav2} and have come out negatively. Limits on the abundance of anomalous gold isotopes range from $6\times 10^{-12}$ for light isotopes $\sim 3$ GeV heavier than a gold nucleus to $1\times 10^{-8}$ for heavy isotopes up to 1480 GeV heavier than a gold nucleus. In \cite{jav} these limits were used to constrain the contribution of SIMPs to the total energy density of the Universe as a function of the SIMP mass.

Currently, there are about 20 dark matter direct detection experiments underway. In these experiments one tries to detect recoils of nuclei caused by collisions with DM particles. Most of these experiments use sophisticated background discrimination techniques to try to extract a DM signal. An alternative way to extract a DM signal from the background is to measure the annual modulation in the dark matter flux caused by the Earth's rotation around the Sun \cite{mod,calc}. The DAMA/NaI experiment \cite{dama} has used this technique and is the only one to have reported a positive signal. The results of this experiment are controversial because other more sensitive searches have not detected nuclear recoils due to DM interactions. For WIMPs interacting with nuclei through spin independent interactions, the available parameter space is very limited \cite{gelm}. The case of spin dependent interactions is studied in \cite{sav}. Inelastic dark matter \cite{weiner} and mirror matter \cite{foot} have also been proposed as an explanation for the DAMA results.

In this article we consider a scenario in which SIMPs make out a small fraction of the dark matter density of the Universe. We assume that the remaining part of the dark matter consists of SIDM particles with almost no interactions with nuclei but with strong interactions, similar in magnitude to their self-interactions, with SIMPs. We shall show that dark matter direct detection experiments can detect signals from a SIDM particle colliding with a SIMP trapped in a nucleus and that this effect may be responsible for the annual modulation signal observed in the DAMA experiment. Since in this model the self-interactions of the SIDM particles don't matter, we shall call them WIMPs and refer to this model as the WIMP-SIMP model.

The DAMA/NaI set up consists of 9 scintillating thallium doped NaI crystals of 9.7 kg each. Electrons and nuclei recoiling after a collision cause emissions of photons that are detected using photomultiplier tubes. Nuclear recoils give rise to a weaker photon signal than electron recoils. The ratio of the two signal strengths is called the quenching factor and is 0.3 for sodium recoils and 0.09 for iodine recoils. The measured signals are expressed in electron equivalent recoil energies, i.e.\ the energy of a recoiling electron needed to account for the observed signal. After a total exposure of 107731 kg.day an annual modulation in the energy range 2-6 KeV was observed with an amplitude of $0.0192\pm 0.0031$ counts/day/kg/KeV \cite{dama}.

In a WIMP-SIMP scenario, assuming that SIMPs don't bind to sodium, all recoils contributing to the annual modulation should come from anomalous iodine isotopes. When a WIMP scatters off a nucleus, the nucleus recoils with an energy, $E_{\text{recoil}}$, of:
\begin{equation}\label{rec}
E_{\text{recoil}}=\frac{\mu^{2}_{\text{WIMP},\text{nucleus}} v^{2}}{M_{\text{nucleus}}}\haak{1-\cos\haak{\theta}},
\end{equation}
where $\theta$ is the scattering angle in the center of mass frame, $v$ the velocity of the WIMP and $\mu_{\text{WIMP},\text{nucleus}}$ is the reduced mass of the WIMP-nucleus system. The DAMA data have been analyzed assuming different types of interactions of nuclei with WIMPs and a wide variety of WIMP halo models \cite{dama}. For our purpose, the case of spin independent interactions is the most interesting. In this case the WIMP-nucleus cross section scales with the square of the atomic mass number and with the square of the relative WIMP-nucleus mass. This means that most collisions are with iodine nuclei. Instead of exploring the full window allowed by the DAMA data, we'll use that a WIMP mass of 50 GeV is consistent with the DAMA data for spin independent interactions. For $M_{\text{WIMP}}\approx 50$ GeV the factor $\haak{\mu_{\text{WIMP},\text{nucleus}}}^{2}/M_{\text{nucleus}}$ in \eqref{rec} is approximately 11 GeV. If we assume that the quenching factor for anomalous iodine isotopes is the same as for ordinary iodine, then constraining the WIMP and SIMP masses such that:
\begin{equation}\label{cnd3}
\frac{\mu^{2}_{\text{WIMP},\text{I}+\text{SIMP}}}{M_{\text{SIMP}}+M_{\text{I}}}= 11\text{ GeV},
\end{equation}
where $\mu_{\text{WIMP},\text{I}+\text{SIMP}}$ is the reduced mass of the WIMP-anomalous iodine system and $M_{\text{I}}$ is the mass of an iodine nucleus, will guarantee that the WIMP-SIMP model is consistent with the DAMA data.

We have calculated the WIMP-SIMP cross section in terms of the abundance of anomalous iodine isotopes by assuming an isothermal halo model \cite{mod} for WIMPs and the value of 11 GeV for the factor $\haak{\mu_{\text{WIMP},\text{nucleus}}}^{2}/M_{\text{nucleus}}$ in \eqref{rec}. We have assumed that the local WIMP density is 0.3 GeV/cm$^{3}$, the WIMP velocity dispersion is 270 km/s and assumed that the escape velocity is infinite. By using that the Sun moves around the center of the galaxy with a velocity of 232 km/s and that the Earth rotates around the Sun with a velocity of 30 km/s at 60 degrees inclination with respect to the galactic plane, we have found that the DAMA annual modulation amplitude implies (see \cite{calc, gelm2} for details about such calculations):
\begin{equation}
\sigma_{\text{WIMP}-\text{SIMP}}=\haak{0.29\pm 0.06} \frac{m_{\text{WIMP}}}{\text{GeV}}\frac{1}{f_{\text{SIMP}}} \text{ pico-barns}.
\end{equation}
Here $\sigma_{\text{WIMP}-\text{SIMP}}$ is the WIMP-SIMP cross section, $f_{\text{SIMP}}$ is the abundance of anomalous iodine nuclei. If $\sigma_{\text{WIMP}-\text{SIMP}}$ is equal to $\sigma_{\text{dd}}$ in Eqn.\ \eqref{range}, then:
\begin{equation}\label{rng}
2.5\times 10^{-14}\lesssim f_{\text{SIMP}} \lesssim 4.5\times 10^{-13}.
\end{equation}
This is less than the best limit of $6\times 10^{-12}$ for light anomalous gold isotopes and far less than the limit of $1\times 10^{-8}$ for heavy gold isotopes \cite{jav2}. Of course, the lower the abundance of SIMPs, the higher $\sigma_{\text{WIMP}-\text{SIMP}}$ can be before SIMPs would interfere with galactic structure formation. So, the assumption $\sigma_{\text{WIMP}-\text{SIMP}}=\sigma_{\text{dd}}$ leading to \eqref{rng} is very conservative.

Most other direct detection experiments use elements that are lighter than iodine. The CDMS II experiment \cite{cdms} uses germanium and the null results of this experiments pose the strongest constraints on spin independent WIMP-nucleon cross sections. The null results of CDMS II and the annual modulation observed by the DAMA experiment can be simultaneously explained if SIMPs don't bind to germanium or lighter elements and if the WIMP-nucleon interaction is sufficiently weak. The WIMP-nucleus cross section caused by the exchange of virtual SIMPs, $\sigma_{w,n}$, should be of the order:
\begin{equation}\label{est}
\sigma_{w,n}\sim \frac{\mu_{\text{WIMP},\text{nucleus}}^{2}}{M_{\text{SIMP}}^{4}}A_{\text{nucleus}}^{2}g^{2}.
\end{equation}
Here $\mu_{\text{WIMP},\text{nucleus}}$ is the relative mass of the WIMP-nucleus system, $A_{\text{nucleus}}$ is the atomic mass number of the nucleus and $g$ is the coupling between SIMPs and nucleons. According to the results of the CDMS II experiment, spin independent WIMP-nucleon cross sections larger than $4\times 10^{-43}$ cm$^2$ are excluded at 90\% confidence level \cite{cdms}. This corresponds to $2\times 10^{-36}$ cm$^2$ for the WIMP-germanium cross section. We shall constrain the WIMP-SIMP model by demanding that $\sigma_{w,n}\leq 1$ pico-barn for germanium. According to \eqref{est} this implies:
\begin{equation}\label{cnd1}
\haak{\frac{\mu_{\text{WIMP},\text{germ}}}{M_{\text{SIMP}}}}\frac{\text{GeV}}{M_{\text{SIMP}}}=6.8\times 10^{-7}\frac{x}{g}.
\end{equation}
Here $\mu_{\text{WIMP},\text{germ}}$ is the relative mass of the WIMP-germanium system and $x\leq 1$ parametrizes the available window. Demanding that SIMPs don't bind to germanium nuclei constrains the coupling $g$. Using the assumption that SIMPs interact with nuclei in a similar way as hyperons, a relation between the minimum atomic mass number, $A_{\text{min}}$, below which SIMPs don't bind to nuclei, and the SIMP-nucleon cross section was obtained in \cite{tep}. If we
assume that the SIMP-nucleon cross section is $g^{2}$ barnes, this relation implies:
\begin{equation}
\frac{g}{2\text{ GeV}}A_{\text{min}}^{2/3}=\frac{1}{M_{\text{SIMP}}}+\frac{1}{M_{A_\text{min}}}.
\end{equation}
If we take $A_{\text{min}}\approx 100$, $g$ becomes:
\begin{equation}\label{cnd2}
g=0.1\haak{10^{-2}+\frac{\text{GeV}}{M_{\text{SIMP}}}}.
\end{equation}

\begin{figure}
\setlength{\unitlength}{0.07 \textwidth}
\begin{center}

\begin{picture}(5,5)
\put(2.5,1.5){\makebox(0,0){\includegraphics[width=0.45\textwidth]{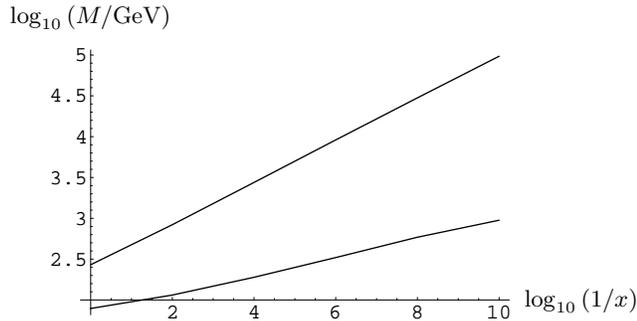}}}
\put(0,3.1){\makebox(0,0){$\log_{10}\haak{M/\text{GeV}} $}}
\put(5.2,0.1){\makebox(0,0){$\log_{10}\haak{1/x}$}}
\end{picture}
\caption{The upper and lower curves are $M_{\text{SIMP}}$ and $M_{\text{WIMP}}$ as a function of $\log_{10}(1/x)$, respectively.}\label{res}
\end{center}
\setlength{\unitlength}{1 pt}
\end{figure}

Using the conditions \eqref{cnd1}, \eqref{cnd2} and \eqref{cnd3}, $M_{\text{WIMP}}$ and $M_{\text{SIMP}}$ can be calculated as a function of the parameter $x$ defined in Eq.\ \ref{cnd1}, see Fig.\ \ref{res}.

E.g., if $x=10^{-2}$, a WIMP detector could detect nuclear recoils in germanium consistent with a
spin independent WIMP-nucleon cross sections of $2\times 10^{-45}$ cm$^2$ for a WIMP mass of about
50 GeV, apparently proving that the DAMA/NaI results are wrong. However, according to the WIMP-SIMP model and the assumptions we have made here, this would still be consistent with the DAMA results if $M_{\text{WIMP}}=110$ GeV and $M_{\text{SIMP}}= 850$ GeV.

Three other experiments are using elements heavier than iodine and have not detected nuclear recoils due to DM interactions. The CRESST II experiment \cite{cresst} uses CaWO$_4$ and has a lower nuclear lower nuclear recoil detection threshold, 12 KeV for tungsten compared to 22 KeV for iodine in the DAMA/NaI experiment. No recoil events in tungsten were detected during a 20.5 Kg day exposure. We have calculated that if the abundance of anomalous tungsten isotopes in the CRESST II detector were the same as the abundance of anomalous iodine isotopes in the DAMA/NaI detector, there should have been about 17 events. This means that the abundance of anomalous tungsten isotopes should be at least an order of magnitude less than the abundance of anomalous iodine isotopes. Because SIMPs don't penetrate a large distance below the surface, the abundance of anomalous isotopes will strongly depend on the geological history of the material. Samples of tungsten and iodine typically have different geological histories \cite{geo}. Tungsten originates from deep within the Earth. All tungsten deposits are of magmatic or hydro thermal origin. Iodine, on the other hand, is usually found in salt mines. The minerals found in salt mines were originally dissolved in a sea. The salt mine formed when that sea receded and dried out. The SIMP-nucleon cross section implied by \eqref{cnd2} implies a stopping distance of the order of tens of kilometers \cite{tep}. So, unless the tungsten used in the CRESST II experiment came from very old deposits, the iodine in the DAMA/NaI detectors should have a larger abundance of anomalous isotopes than the tungsten used in the CRESST II experiment. Unfortunately, the CRESST II collaboration hasn't got the information about the location where their tungsten was obtained from \cite{coz}.

The DAMA/Xe-2 \cite{bern} and the ZEPLIN I \cite{zeplin} experiments use liquid xenon. Since xenon has roughly the same atomic mass number as iodine, the same abundance of anomalously heavy isotopes would lead to the same apparent spin independent WIMP-nucleon cross section. In the DAMA/Xe-2 experiment xenon enriched in $^{129}$Xe was used. No events were detected and this was consistent with the results of the DAMA/NaI experiment assuming spin independent WIMP interactions.
The ZEPLIN I experiment used natural xenon. The 90\% exlusion limit from this experiment for spin independent WIMP-nucleon cross sections is about $1\times 10^{-6}$ pico-barns, about a factor of 2 below the lower limit of the DAMA/NaI experiment. Assuming that the WIMP-SIMP model is correct, this means that the abundance of anomalous xenon isotopes in the ZEPLIN detector is less than the abundance of anomalous iodine isotopes in the DAMA/NaI detector. Like in case of tungsten, this could be due to a geological effect. There is some evidence that heavy noble gasses in the atmosphere are recycled in the mantle \cite{xen}. It is thus possible that (very) heavy isotopes of xenon in the atmosphere are depleted by such a process.

In the near future, the WIMP-SIMP model presented here can be verified or ruled out using data from dark matter direct detection experiments that are currently underway. If experiments using different materials report results that are consistent with a single WIMP model, then that would rule out the WIMP-SIMP model. If, on the other hand, only experiments using elements lighter than some critical atomic mass number are consistent with each other while experiments using heavier elements detect signals that are too large to be explained using conventional WIMP models, then that would be consistent with the WIMP-SIMP model. Experiments that use xenon are especially interesting. Using isotope separation methods to enhance to concentration of lighter xenon isotopes one could deplete the concentration of anomalously heavy isotopes, thus influencing the WIMP detection rate. Since xenon has 9 stable isotopes of which two have an odd atomic mass number this effect should be distinguishable from the predictions of any conventional WIMP model.

In conclusion, we have shown that the annual modulation signal detected by DAMA can be explained by WIMPs interacting with SIMPs captured in iodine nuclei. The WIMP-SIMP cross section can be in the same range as is postulated for self-interacting dark matter without violating experimental bounds on anomalously heavy isotopes.

\begin{acknowledgments}
I thank David Sinclair for providing information on the geological origin of iodine and tungsten, Cristina Cozzini for helpful conversations on the CRESST II experiment, John Arabadjis for references on recent SIDM simulations and Vitaly Kudryavtsev for answering questions about the Zeplin I experiment. 
\end{acknowledgments}

\end{document}